%% file: main.tex
\newif\ifSingleColumn
\SingleColumnfalse
\ifSingleColumn
\documentclass[draftclsnofoot, onecolumn, 12pt]{IEEEtran}
\else
\documentclass[twocolumn,journal,10pt]{IEEEtran}
\fi

\usepackage[draft]{hyperref}

\input{header.tex}

\input{defines.tex}

\begin{document}

\title{\linespread{1.2}\huge{Distillation-Enabled Knowledge Alignment Protocol for Semantic Communication in AI Agent Networks}}

\author{\linespread{1.25}
\IEEEauthorblockN{
\normalsize{Jingzhi~Hu},~\IEEEmembership{\normalsize Member,~IEEE} and
\normalsize{Geoffrey Ye Li},~\IEEEmembership{\normalsize Fellow,~IEEE} 
}
\thanks{This work has been submitted to the IEEE for possible publication. Copyright may be transferred without notice, after which this version may no longer be accessible. J. Hu and G. Y. Li are with the Department of Electrical and Electronic Engineering, Imperial College London, London SW7 2AZ, UK.~(email: \{jingzhi.hu, geoffrey.li\}@imperial.ac.uk)}
\thanks{Code is available at \url{https://github.com/DJ-Duke/DeKAP}}
}

\maketitle
\begin{abstract}
\input{./sections/1_abstract.tex}

\end{abstract}

\ifSingleColumn
\newpage
\fi

\input{sections/2_introduction.tex}

\input{sections/3_system_model.tex}

\input{sections/4_perform_analysis.tex}

\input{sections/5_evaluation.tex}

\input{sections/6_conclusion.tex}

\renewcommand{\refname}{References} 

\bibliographystyle{IEEEtran}
\bibliography{bibilio}

\end{document}

%% file: header.tex
\usepackage[utf8]{inputenc}
\usepackage{glossaries}
\usepackage[symbols,nogroupskip]{glossaries-extra}
\usepackage{bm}
\usepackage{amssymb}
\usepackage{amsmath}
\usepackage{amsfonts}
\usepackage{mathrsfs}
\usepackage{bbm}
\usepackage{scalerel} 
\usepackage{graphicx}
\usepackage{cite}
\usepackage{enumerate}
\usepackage{url}
\usepackage{epstopdf}
\usepackage{algorithm}
\usepackage[noend]{algpseudocode}
\usepackage{algorithmicx,algorithm}
\usepackage{float}
\usepackage{amsthm}
\usepackage{color}
\usepackage{enumitem}
\usepackage{clipboard}
\usepackage[normalem]{ulem} 
\usepackage{framed}
\usepackage{array} 
\usepackage{hhline}
\usepackage[table]{xcolor}
\usepackage{balance}
\usepackage{subcaption}
\usepackage{xargs} 
\usepackage{cooltooltips}
\usepackage{xcolor}  
\usepackage{comment}

\newcommand{\fref}[2]{Fig.~\ref{#1}(\subref*{#2})}

\newcommand{\beq}{\begin{equation}}
\newcommand{\eeq}{\end{equation}}
\newcommand{\beqq}{\begin{equation*}}
\newcommand{\eeqq}{\end{equation*}}
\newcommand{\beaq}{\begin{align}}
\newcommand{\eeaq}{\end{align}}

\DeclareMathOperator*{\argmin}{argmin}

\newtheorem{proposition}{\bf Proposition}

\allowdisplaybreaks

\newcommand{\rev}{\textcolor[rgb]{0,0,0}}

%% file: defines.tex
\newcommand{\vGama}{\varGamma}

\newcommand{\bmX}{\bm X}
\newcommand{\bmY}{\bm Y}

\newcommand{\bmB}{\bm B}
\newcommand{\bmA}{\bm A}
\newcommand{\bmP}{\bm P}

\newcommand{\bmf}{\bm f}

\newcommand{\cX}{\mathcal X}
\newcommand{\cY}{\mathcal Y}
\newcommand{\cZ}{\mathcal Z}
\newcommand{\cD}{\mathcal D}

\newcommand{\cK}{\mathcal K}

\newcommand{\cP}{\mathcal P}

\newcommand{\cN}{\mathcal N}
\newcommand{\cM}{\mathcal M}
\newcommand{\cL}{\mathcal L}

\newcommand{\rT}{\mathrm T}

\newcommand{\rA}{\mathrm A} 
\newcommand{\rS}{\mathrm S}

\newcommand{\net}{\mathrm{net}}

\newcommand{\pt}{\mathrm{pt}}
\newcommand{\enc}{\mathrm{E}}
\newcommand{\dec}{\mathrm{D}}

\newcommand{\figwidth}{0.33\linewidth}

%% file: sections/1_abstract.tex
Future networks are envisioned to connect massive artificial intelligence~(AI) agents, enabling their extensive collaboration on diverse tasks.
Compared to traditional entities, \rev{these} agents naturally suit the semantic communication~(SC), which can significantly enhance the bandwidth efficiency.
Nevertheless, SC requires \rev{the} knowledge \rev{among} agents \rev{to be aligned}, while agents have distinct expert knowledge for their individual tasks \rev{in practice}. 
In this paper, we propose \rev{a} distillation-enabled knowledge alignment protocol~(DeKAP),
\rev{which} distills the expert knowledge of each agent into parameter-efficient low-rank matrices, allocates them across the network, \rev{and allows} agents to simultaneously maintain aligned knowledge for multiple tasks.
We formulate the joint minimization of alignment loss, communication overhead, and storage cost as a large-scale integer linear programming problem and \rev{develop} a highly efficient greedy algorithm.
\rev{From computer simulation, the} DeKAP establishes \rev{knowledge alignment} with the lowest communication and computation resources compared to conventional approaches.

%% file: sections/2_introduction.tex
%%%%%%%%%%
\section[Introduction]{Introduction}
%%%%%%%%%%
\label{SEC_INTRO}

\rev{Future communication} networks will usher in a new era of the “Internet of Intelligence,” where \rev{human beings, devices, and} a wide range of artificial intelligence~(AI) agents are seamlessly interconnected~\cite{Chen20SUMMIT_From}. 
Through communications, distributed agents can collaborate on tasks by sharing their data and results, extending their service range and coverage.
However, intensive collaboration among agents will generate a substantial traffic load, exacerbating the heavy burden upon the already crowded network.

Fortunately, as AI agents are equipped with powerful neural models for feature encoding and decoding, they naturally suit the bandwidth-efficient semantic communication~(SC) paradigm~\cite{Xie21TSP_Deep}.
The fundamental principle of SC is to transmit only the semantic features of data closely pertaining to the completion of a target task.
In an SC link between agents, the transmitter~(Tx) agent first encodes the high-dimensional data into semantic features using its neural model \rev{and sends them to the} receiver~(Rx) agent, \rev{which then} decodes them into task results.
Since the task completion often requires only a small portion of information in the data, the semantic features are expected to have a much smaller size than the \rev{original} data.
Therefore, SC \rev{can} significantly reduce the traffic load.

While SC is promising, its effectiveness relies on a critical prerequisite: The~Tx and Rx agents must have \emph{aligned knowledge} \rev{on} the target task \rev{for semantic encoding and decoding}.
Most studies \rev{on} SC assume perfectly aligned knowledge in Tx and Rx~\cite{Qin24PIEEE_AI,Bourtsoulatze19TCCN_Deep}. 
However, agents in practice may develop distinct expert knowledge during the deployment for their individual application tasks~\cite{Rossi22Huawei_Network}.
\rev{Therefore, the knowledge alignment (KA) turns an important issue in SC.}

Existing approaches for the KA can be categorized into \emph{adaptation-based} and \emph{equalization-based}.
The first category enables the Tx and Rx to mutually adapt their neural parameters.
In~\cite{Zhang23JSAC_Deep}, an Rx-lead training process is proposed, where the Rx feedbacks gradients to adapt the encoder of the Tx. 
\rev{The} latent-space-based adapting method \rev{in~\cite{Si24TWC_Post} uses} semantic features to adapt the parameters.
\rev{In~\cite{Choi24TVT_Semantics},} the Tx downloads the neural model of the Rx to adapt locally.
\rev{The other category of} approaches equalize the latent spaces of semantic features of Tx and Rx by transformation.
\rev{In~\cite{Sana23GC_Semantic}}, the latent spaces \rev{are divided} into atomic subspaces, and the transformations \rev{are optimized among these} subspaces.
\rev{In~\cite{Fiorellino24Arxiv_Dynamic},} the correlation between data and a set of anchor data \rev{is leveraged} to construct encoder-equivalent semantic features.
Furthermore, \rev{in~\cite{Hu24Arxiv_Zero}} continual learning \rev{is used} to preserve equivalent latent spaces when agents develop discrepant expert knowledge.
 
However, the existing approaches only enable Tx and Rx to achieve alignment on a single task-specific knowledge.
In stark contrast, a network of AI agents may develop multiple expert knowledge due to \rev{their} diverse individual tasks.
Ideally, network-level agent collaboration should allow each agent to leverage data from any other agent and perform every task for which expert knowledge is available.
Therefore, for SC to fully boost agent collaboration, \rev{it} is imperative \rev{to develop a novel KA approach to} maintain multiple aligned expert knowledge simultaneously.

\rev{In this article}, we propose a novel \textbf{d}istillation-\textbf{e}nabled \textbf{k}nowledge \textbf{a}lignment \textbf{p}rotocol~(DeKAP) at a network scale for multiple expert knowledge.
Inspired by the parameter-efficient fine-tuning techniques \rev{in}~\cite{Hu22ICLR_LoRA}, \rev{the} DeKAP converts each expert knowledge into distilled knowledge~(DK) of a significantly smaller size, e.g., with a parameter ratio~(PR) as low as $1\%$\rev{, which can} be efficiently shared and stored by \rev{all agents}.
To optimize the DeKAP, we \rev{first} analyze its alignment loss, communication overhead, and storage cost, and \rev{find} that their joint minimization is a large-scale integer linear programming~(ILP) problem.
To tackle the prohibitive complexity of the ILP, we propose a highly efficient greedy algorithm, \rev{which} achieves near-optimal performance with significantly lower complexity than the state-of-the-art ILP solver.

The rest of this paper is organized as follows. 
In Sec.~\ref{sec_system_lodel}, we model AI agent networks and the SC between AI agents.
In Sec.~\ref{SEC_II_D}, we propose the DeKAP and optimize its performance.
Evaluation setups and results are presented in Sec.~\ref{SEC_EVAL}, and a conclusion is drawn in Sec.~\ref{SEC_CONCLU}.

%% file: sections/3_system_model.tex
%%%%%%%%%%
\section[System Model]{System Model}
%%%%%%%%%%
\label{sec_system_lodel}

In this section, we establish a general model for AI agent networks and SC between AI agents.

%========================================
\subsection{AI Agent Networks} \label{SEC_II_A} 
%========================================

\rev{As shown in Figure~\ref{fig_sys_lod}, an} AI agent network operates on top of the infrastructure of a set $\cN$ of $N$ nodes with computational resources and storage spaces.
Each pair of nodes can communicate via reliable links enabled by Wi-Fi, cellular, and/or wired connections, depending on the infrastructure.
Each node has a local AI agent implemented as a deep neural model, which is essentially a parameterized function processing input data into its corresponding result for a certain task.

In particular, we focus on the neural models of agents that follow the encoder-decoder architecture for image processing tasks, e.g., noise removal, super resolution, anomaly detection, etc.
The input data and output result of the neural model are represented by $\bmX\in\cX$ and $\bmY\in\cY$, respectively, with $\cX$ and $\cY$ being their high-dimensional value spaces.
Then, the neural model can be represented by $\bmf=(\bmf^{\enc},\bmf^{\dec})$, with the image processing being expressed as,
\beq
\bmY = \bmf(\bmX;\cP) = \bmf^{\dec}\big(\bmf^{\enc}(\bm X;\cP^{\enc});\cP^{\dec}\big),
\eeq
\rev{where} $\bmf^{\enc}\!: \cX\rightarrow\cZ$ and $\bmf^{\dec}\!:\cZ\rightarrow\cY$ are the functions of the encoder and decoder with $\cZ$ denoting the feature space; $\cP$, $\cP^{\enc}$, and $\cP^{\dec}$ are the parameter sets for the complete neural model, the encoder, and the decoder, respectively. 
As the parameter sets determine the mapping from input data to output results, they contain the knowledge of the neural model.

Following the common practices of AI deployment, we assume that the neural models of all the agents initially have the same pre-trained parameters, $\cP_{\pt}$.
During their deployment, agents become specialized in their individual tasks and develop distinct expert knowledge given the locality of both application preferences and training data availability. 
For agent $i$~($i\in\cN$), its individual task is represented by a joint probability distribution of data and results, i.e, $\vGama_i: (\cX,\cY)\rightarrow [0,1]$.
Being specialized in $\vGama_i$ changes the parameters of agent $i$ from $\cP_{\pt}$ to $\cP_{i}$.
The change $\Delta\cP_i = \cP_{i} - \cP_{\pt}$ represents the developed expert knowledge, which is generally obtained by
\beq
\label{equ_expert_know}
\Delta\cP_i=\argmin_{\Delta \cP'_i}~\mathop{\mathbb E}\limits_{(\bmX,\bmY)\sim\varGamma_i} \Big(J_i(\bmf(\bmX; \cP_{\pt}+\Delta\cP'_i),\bmY) \Big),
\eeq
where $J_i(\bmY',\bmY)$ is the loss function of agent $i$'s task.

Agents with distinct expert knowledge and data access can collaborate.
The ideal collaboration should allow each agent to accomplish any tasks for which \rev{others} in the network have available data and expert knowledge.
As shown in \fref{fig_sys_lod}{fig_sys_lod_a}, when agent $j$ needs the task result for which agent $k$ owns the knowledge while agent $i$ owns the data, the data is firstly sent from \rev{agent} $i$ to $k$, then the task result is sent to agent $j$.

\ifSingleColumn
    \renewcommand{\figwidth}{0.35\linewidth}
\else
    \renewcommand{\figwidth}{0.48\linewidth}
\fi

\begin{figure}[t]
    \centering 
    \begin{subfigure}{\figwidth}
        \centering
        \includegraphics[width=\linewidth]{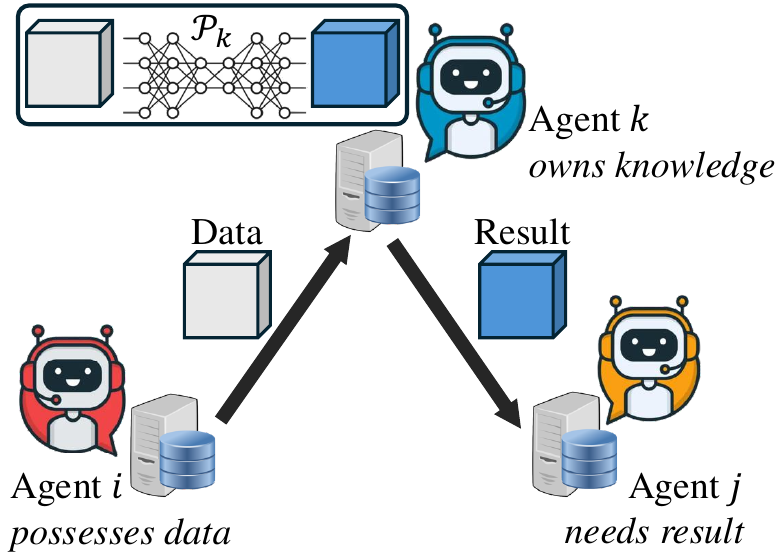}  
        \caption{}
        \label{fig_sys_lod_a}
    \end{subfigure}
    \ifSingleColumn
    \hspace{3ex}
    \fi
    \begin{subfigure}{\figwidth}
        \centering
        \includegraphics[width=\linewidth]{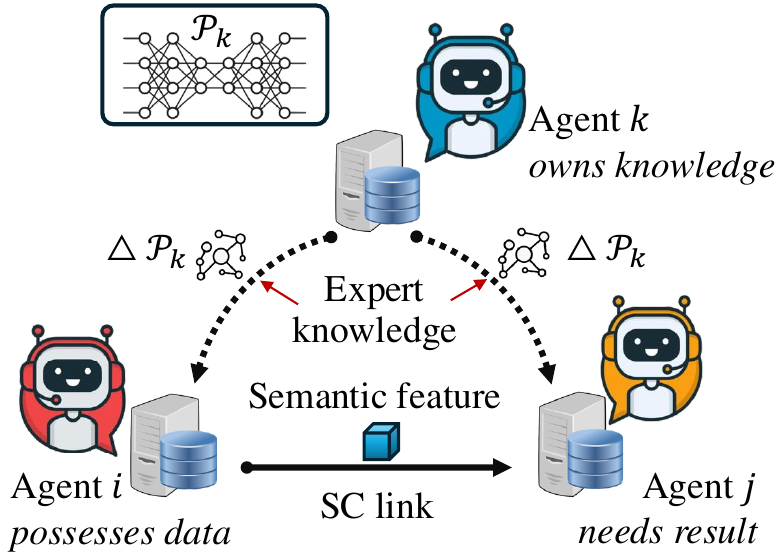}  
        \caption{}
        \label{fig_sys_lod_b}
    \end{subfigure}
    \caption{Collaboration in AI agent networks (a) without SC and (b) with SC and shared expert knowledge.}
    \label{fig_sys_lod}
\end{figure} 

%========================================
\subsection{SC between AI Agents}\label{s2ec_sc_ai}
%========================================

\rev{Intuitively,} the collaboration scheme in \fref{fig_sys_lod}{fig_sys_lod_a} is inefficient because it results in heavy traffic caused by high-dimensional data and result transmission.
To reduce the traffic load, it is promising to exploit the intrinsic encoding and decoding capability of AI agents, enabling data and results to be represented by low-dimensional semantic features.
Then, in accordance with the SC paradigm, agents can communicate via the semantic features instead of the original data and results. 

However, agents in the network may possess distinct expert knowledge in their respective encoders and decoders.
Since the expert knowledge of different agents is not obtained by joint training, they can be significantly misaligned, thus resulting in substantial alignment loss in SC.
We define the \emph{alignment loss} of the SC link from agent $i$ to $j$ for agent $k$'s task as the difference between the results of the SC and those of agent $k$, measured by the task loss function.
It can be calculated by
\beq
\label{equ_align_loss}
J_{\rA,ijk} \!=\! \mathop{\mathbb E}\limits_{\bmX\sim\varGamma_{\bmX,k}} 
J_k\big(\bmf^{\dec}(\bmf^{\enc}(\bmX;\cP^{\enc}_i);\cP^{\dec}_j) , \bmf(\bmX;\cP_k)\big),
\eeq
where $\varGamma_{\bmX,k}$ is the data marginal distribution of agent $k$'s task.

To mitigate the alignment loss, sharing the expert knowledge of agent $k$ is the most intuitive approach, as illustrated in \fref{fig_sys_lod}{fig_sys_lod_b}. 
However, since the expert knowledge is intricately embedded in the neural parameters, sharing it requires transmitting the entire parameter set, leading to substantial communication overhead as well as storage cost.

%%%%%%%%%%%%%%%%%%%%%%%%%%%%%%%%%%%%%%%%%
\section{Distillation-Enabled Knowledge Alignment Protocol}\label{SEC_II_D} 
%%%%%%%%%%%%%%%%%%%%%%%%%%%%%%%%%%%%%%%%%

\rev{In this section, we} propose DeKAP to achieve KA in AI agent networks.
Inspired by the parameter-efficient fine-tuning techniques \rev{in}~\cite{Hu22ICLR_LoRA} and self-knowledge distillation \rev{in}~\cite{zhang2021self}, \rev{our} DeKAP generates DK from each expert knowledge so that agents can efficiently share and store aligned knowledge for various tasks.
The protocol comprises two parts: the \emph{distillation} of expert knowledge and the \emph{allocation} of DK.

%========================================
\subsection{Distillation of Expert Knowledge}
%========================================
\rev{The} DeKAP distills expert knowledge into a set of low-rank matrices to reduce the number of parameters.
Below, we consider an arbitrary agent and omit its index.
Assume that parameter set $\Delta\cP$ of an agent's expert knowledge comprises $M$ parameter matrices, one for each layer in the neural model. 
Denote the $m$-th matrix by $\bmP_{m}\in\mathbb R^{I_{m}\times O_{m}}$, where $I_{m}$ and $O_{m}$ are the dimensions of input and output of the $m$-th layer, respectively.
The DK for $\Delta\cP$ comprises $M$ pairs of rank-$r$ matrices, which can be expressed as,
\beq
\cD=\{\bmB_m\bmA_m\mid \bmB_m \!\in\! \mathbb R^{I_{m}\times r},\bmA_m \!\in\! \mathbb R^{r\times O_{m}}, m\in\cM  \},
\eeq
where $\cM=\{1,...,M\}$.
Given that $r$ is much smaller than $I_{m}$ and $O_{m}$, the parameter ratio~(PR) between the DK and $\Delta\cP$ can be very small, which is calculated by
\beq
\gamma = \sum_{m\in\cM} {r\times (I_{m} + O_{m})}/({I_{m}\times O_{m}}).
\eeq

In the distillation, the agent optimizes $\cD$ for the minimization of the task loss caused by the difference between the results of the DK and that of the expert knowledge, i.e.,
\beq
\label{equ_know_distill}
\min_{\cD}~J_{\rA}(\cD) = \hspace{-.3em}\mathop{\mathbb E}\limits_{\bmX\sim\varGamma_{\bmX}}\hspace{-.3em} 
    J\big(\bmf(\bmX;\cP_{\pt}\!+\!\cD) , \bmf(\bmX;\cP_{\pt}\!+\!\Delta\cP)\big).
\eeq
We note that the objective in~\eqref{equ_know_distill} is the same as the alignment loss in~\eqref{equ_align_loss}.
Therefore, if two agents have the optimized DK $\cD$, the alignment loss of their SC for the task will be minimized.

Given a larger PR, higher ranks can be assigned to the matrices in $\cD$, increasing its potential to achieve a lower loss in~\eqref{equ_know_distill}. 
Nevertheless, it also raises the cost of sharing and storing the DK.
The tradeoff may vary across agents, depending on their frequency of using the DK for SC.
In view of this, DeKAP allows each agent to prepare $L$ levels of DK with increasing PRs, denoted by $\gamma_{1},...,\gamma_{L}$, so that agents can be allocated different levels of DK.
Moreover, every low-rank matrix of a lower-level DK should be a sub-matrix of the corresponding one of a higher-level DK, thereby ensuring that different levels of DK are mutually compatible.

Distillation of multi-level DK requires modification to~\eqref{equ_know_distill} as the alignment loss of each level of DK should be minimized.
Instead of jointly minimizing the alignment loss for all the $L$ levels, we randomly alternate between optimizing different levels of DK.
More specifically, in the $t$-th iteration of distillation, the update of the DK is calculated by
\beq
\cD^{(t+1)} = \cD^{(t)} - \alpha\nabla_{\cD} J(\cD^{(t)}[l^{(t)}]),
\eeq
where $\alpha$ is the step size, $\cD[l]$ denotes the DK of level $l\!\in\!\cL$ with $\cD\!=\!\cD[L]$, and $l^{(t)}$ is randomly sampled in $\cL=\{1,\dots L\}$.

Furthermore, since not all agents' individual tasks \rev{are required} by others, \rev{we denote $\cK\subseteq\cN$ as the set of the agents' tasks whose expert knowledge is needed across the network.}
Consequently, we can represent the generated DK for all the tasks in $\cK$ and levels in $\cL$ by $\cD[k,l]$ ($\forall k\in\cK,l\in\cL$).

One of the most prominent advantages of the DeKAP is that the KA for multiple tasks is maintained simultaneously by the DK.
Each DK can be added, modified, or removed without affecting the others. 
In this regard, the DeKAP converts the intricate KA problem into an allocation problem of~DK.

%% file: sections/4_perform_analysis.tex
\subsection{Allocation of Distilled Knowledge}

We \rev{first} analyze the performance of DeKAP in three aspects, \rev{alignment loss, transmission overhead, and storage cost}:

\textit{\textbf{Alignment Loss}}:
We first \rev{look} the statistical performance of the DeKAP. 
\rev{Denote $F_{ij}[k]$ as} the frequency that agent $i$ as the Tx performs SC with agent $j$ as the Rx for task $k$.
We refer to the selection of different levels of DK as the \emph{exploitation policy}, denoted by binary variable $e_{ij}[k,l]\in\mathbb B$, with $\sum_{l\in\cL} e_{ij}[k,l]=1$ ($\forall k\in\cK$).
Then, the network-scale alignment loss \rev{can be} calculated by,
\beq
\label{equ_LA}
L_{A} = \sum_{k\in\cK}\sum_{j\neq i} \sum_{l\in\cL} F_{ij}[k]\cdot e_{ij}[k,l]\cdot J_{\rA}[k,l],
\eeq
where $J_{\rA}[k,l]$ is the resulting alignment loss of $\cD[k,l]$ after the distillation, and $\sum_{j\neq i}$ is a notation for $\sum_{i\in\cN}\sum_{j\in\cN,j\neq i}$.

\textit{\textbf{Transmission Overhead}}:
\rev{First, let us find} the overhead for transmitting the DK to ensure alignment in SC when the DK is not locally stored by the agent.
Due to the multi-level characteristic of DK, it can be transmitted in a differential manner.
Specifically, if an agent has $\cD[k,l-1]$ but requires $\cD[k,l]$, only those sub-matrices in $\cD[k,l]$ not included in $\cD[k,l-1]$, denoted by $\Delta\cD[k,l]=\cD[k,l]\setminus \cD[k,l-1]$, need to be transmitted.
In view of this, for the SC link from agent $i$ to $j$ for task $k$, we use \emph{transmission policies} $\phi_{hij}[k,l]\in\mathbb B$ and $\varphi_{hij}[k,l]\in\mathbb B$ ($h\in\cN$) to indicate whether $\Delta\cD[k,l]$ is transmitted from agent $h$ to agents $i$ and $j$, respectively.
Then, the transmission overhead in the network \rev{can be} calculated by
\beq
\label{equ_LT}
\begin{aligned}
O_{\rT} = \sum_{k\in\cK} \sum_{j\neq i} F_{ij}[k]\cdot \sum_{l\in\cL} &\Big(\sum_{h\in\cN}\phi_{hij}[k,l]\cdot T_{hi}[k,l]\\
&+\sum_{h\in\cN}\varphi_{hij}[k,l]\cdot T_{hj}[k,l]\Big),
\end{aligned}
\eeq
\rev{where $T_{hi}[k,l] = S[k,l]/R_{hi}$} represents the overhead for agent $h$ to transmit $\Delta\cD[k,l]$ to $i$, $S[k,l]$ is the size of $\Delta\cD[k,l]$, and $R_{hi}$ is the transmission rate from agent $h$ to $i$.
Moreover, we assume $T_{ii}[k,l]=0$ as self-transmission has no overhead. 

\textit{\textbf{Storage Cost}}: Denote the indicator for agent $i$ to store $\Delta\cD[k,l]$ by $s_{i}[k,l]\in\mathbb B$, which we refer to as the \emph{storage policy}.
The cost for storing $\Delta\cD[k,l]$ is assumed to be $S[k,l]$, and then the total storage cost can be calculated by,
\begin{align}
\label{equ_LS}
C_{\rS} = \sum_{k\in\cK} \sum_{i\in\cN} \sum_{l\in\cL} S[k,l]\cdot s_i[k,l].
\end{align}

Due to the equivalent status of task $k\in\cK$, we hereby focus on an arbitrary task and omit index $k$.
The variables in the allocation of DK comprise the exploitation, transmission, and storage policies.
In addition, we introduce auxiliary variable $\tau_i[l]\in\mathbb B$ to indicate whether level $l$ DK is needed by any SC link of agent $i$.
The complete set of optimization variables is
\beq
\mathcal V \!=\! \{e_{ij}[l],s_i[l],\phi_{hij}[l],\varphi_{hij}[l],\tau_i[l]\mid\forall i,j\!\neq\! i,h,l\},
\eeq
where we adopt short-hand notations for the index ranges, e.g., $\forall i$ means $\forall i\in\cN$, $\forall j\neq i$ means $\forall i,j\in\cN$ and $j\neq i$, etc.

We find that the joint minimization of $L_{\rA}$, $O_{\rT}$, and $C_{\rS}$ can be formulated as an ILP problem,
\begin{subequations}
\label{prob_sp1}
\begin{align}
\label{sp1_cons_0}
\min_{\mathcal V}~&J_{\net} = \eta_{\rA}L_{\rA}+\eta_{\rT}O_{\rT}+\eta_{\rS}C_{\rS},\\
\text{s.t.}~
\label{sp1_cons_1}
& \tau_i[l]\leq \sum_{h\in\cN}\phi_{hij}[l] + s_i[l],\\
\label{sp1_cons_2}
& \tau_j[l]\leq \sum_{h\in\cN}\varphi_{hij}[l] + s_j[l],\\
\label{sp1_cons_3}
& e_{ij}[l']\leq \tau_i[l],~e_{ji}[l']\leq \tau_i[l],\\
\label{sp1_cons_4}
&  \phi_{hij}[l] \leq s_h[l],~\varphi_{hij}[l] \leq s_h[l],\\
\label{sp1_cons_5}
& \sum_{l\in \cL} e_{ij}[l]=1,~\forall i, j\neq i, h, l, l'\geq l.
\end{align}
\end{subequations}
where $J_{\net}$ \rev{denotes} the \emph{network loss}, and $\eta_{\rA},\eta_{\rT},\eta_{\rS}\in[0,1]$ in~\eqref{sp1_cons_0} represent the weights on the performance metrics.
Constraints~\eqref{sp1_cons_1} and~\eqref{sp1_cons_2} ensure that when the SC link from agent $i$ to $j$ uses the DK of level $l$, the DK is available on both sides of the link. 
Constraint~\eqref{sp1_cons_3} is because the DK of a lower level is needed by all the SC links exploiting the DK of a higher level.
Constraint~\eqref{sp1_cons_4} ensures that DK can be transmitted from agent $h$ only if it is stored by $h$.
Finally,~\eqref{sp1_cons_5} ensures that every SC link should exploit DK to achieve KA.

Formulating the allocation of DK as an ILP problem allows the use of the state-of-the-art ILP solver, Gurobi~\cite{gurobi}.
However, it remains challenging due to the massive binary variables and constraints, both of order $\mathcal O(N^3L)$, leading to prohibitive complexity.
To handle this challenge, we propose a highly efficient greedy algorithm based on Proposition~1.

\begin{proposition}
\label{prop_2}
When storage policy $s_i[l]$~($\forall i\in\cN, l\in\cL$) is fixed, the optimal exploitation policies and transmission policies for \eqref{prob_sp1} have the closed-form expressions below:
\begin{align}
\label{equ_e_star}
&e^*_{ij}[l] = \mathbb I\Big(l=\argmin_{l'\in\cL}~\eta_{\rA}J_{\rA}[l']  \\ 
&\hspace{5.5em} + \eta_{\rT}\!\sum_{l''\leq l'}\!\sum_{h\in\cN}\!\big(T_{\min,i}[l'']\!+\!T_{\min,j}[l'']\big)\Big)\!, \nonumber\\
\label{equ_tau_star}
& \tau^*_i[l] = \mathbb I\Big(\sum_{j\in\cN,j\neq i}\sum_{l'\geq l}~e^*_{ij}[l']+e^*_{ji}[l'] \geq 1\Big),\\
\label{equ_phi_star}
&\phi^*_{hij}[l] = \mathbb I(T_{hi}[l]=T_{\min,i}[l])\cdot \mathbb I(\tau^*_{i}[l]=1),\\
\label{equ_vphi_star}
&\varphi^*_{hij}[l] = \mathbb I(T_{hj}[l]=T_{\min,j}[l])\cdot \mathbb I(\tau^*_{j}[l]=1),
\end{align}
\rev{where} $\mathbb I(\cdot)$ is the indicator function, and $T_{\min,i}[l]$ is the minimal transmission overhead to agent $i$, calculated by
\beq
T_{\min,i}[l] = \min_{h\in\cN}~T_{hi}[l]/s_{h}[l].
\eeq
\end{proposition}
\begin{IEEEproof}
Eqn.~\eqref{equ_e_star} is derived from the linearity of~\eqref{sp1_cons_0} with respect to the exploitation policies. Eqn.~\eqref{equ_tau_star} follows from constraint~\eqref{sp1_cons_3}. Eqns.~\eqref{equ_phi_star} and~\eqref{equ_vphi_star} are derived from the minimization of $O_{\rT}$.
\end{IEEEproof}

Based on Proposition~\ref{prop_2}, we propose the highly efficient greedy algorithm for~\eqref{prob_sp1} in Algorithm~\ref{alg_greedy}, which is bound to converge to a local optimum. 

\ifSingleColumn
\begin{figure}[H]
\else
\begin{figure}[H]
\fi
\begin{algorithm}[H]
\ifSingleColumn
\normalsize
\else
\small
\fi
\caption{Greedy algorithm for DK allocation.}
\label{alg_greedy}
\begin{algorithmic} [1]
\State Initialize the storage policy by $s_i[l]=1$~($\forall i\in\cN,l\in\cL$).
\State Fix the storage policies of all the agents except for $i$.
\State Enumerate all the possible storage policies of agent $i$, and select the optimal one for~\eqref{sp1_cons_0} with the help of Proposition~\ref{prop_2}.
\State When all agents have been traversed and agent $i$'s storage policy is unchanged, converge; otherwise, go to step 2 for the next agent.
\end{algorithmic}
\end{algorithm}
\end{figure}

%% file: sections/5_evaluation.tex
%%%%%%%%%%%%%%%%%%%%%%%%%%%%%%%%%
\section{Evaluation}\label{SEC_EVAL}
%%%%%%%%%%%%%%%%%%%%%%%%%%%%%%%%%

In this section, we first describe the experimental setup and then present the evaluation results for the DeKAP.
%========================================
\subsection{Experimental Setup}
%========================================
Each agent has a neural model of the vector-quantized variational auto-encoder~(VQ-VAE)~\cite{Van17NIPS_Neural}.
The VQ-VAE comprises $M=20$ layers with $1.1\times 10^6$ parameters and encodes image data of size $224\times 224$ with three 8-bit color channels into semantic features of $56\times56\times 9$ bits. 
We pre-train the VQ-VAE over the ImageNet dataset for image reconstruction.
Then, the pre-trained parameters are adapted for the agents' individual tasks.
We focus on $K=4$ tasks, including noise removal~(NR), color correction~(CC), super resolution~(SR), and anomaly detection~(AD)\footnote{In the AD task, anomalies are located by the difference between the results of pre-trained parameters and the reconstructed no-anomaly images.}, which are illustrated in Fig.~\ref{fig_eval_1}.

For the distillation of multi-level DK, we adopt $L=5$ PR levels ranging from 1\% to 5\%.
Given a PR, the ranks are uniformly allocated among the layers, and increasing the PR by 1\% averagely raises each rank by $0.74$.
The loss function used in the distillation is the \rev{mean-squared} error~(MSE) between output results combined with their distance in the feature space of VGG16~\cite{vgg16loss}.
The multi-level DK is optimized for 100 epochs per level with step size $\alpha=5\times10^{-4}$.
Each epoch consists of training on 38 batches of 128 images, with performance validated on another set of \rev{1,200} images.

As for the allocation of DK, instead of adopting specific networks, we use random networks with $N\in[3,50]$ agents under normalized conditions.
In~\eqref{equ_LA}, for each agent $i$ and task $k$, frequency $F_{ij}[k]$ ($\forall j\in\cN,j\neq i$) follows a uniform distribution and sums to one.
Every storage cost $S[k,l]$ in~\eqref{equ_LS} is normalized to one. 
Transmission rate $R_{hi}$ below~\eqref{equ_LT} is randomly sampled from the log-normal distribution with zero mean and unit variance.
Finally, the weights in~\eqref{sp1_cons_0} are $\eta_{\rA}=1$, $\eta_{\rT}=0.5$, and $\eta_{\rS}=0.1$.

%========================================
\subsection{Evaluation Results}
%========================================
\label{ssec_eval_res}

\input{sections/eval_res/eval_1}

\input{sections/eval_res/eval_2}

\input{sections/eval_res/eval_3}

%% file: sections/eval_res/eval_1.tex
\ifSingleColumn
    \renewcommand{\figwidth}{0.6\linewidth}
\else
    \renewcommand{\figwidth}{0.9\linewidth}
\fi

\begin{figure}[t] 
    \centering
    \includegraphics[width=\figwidth]{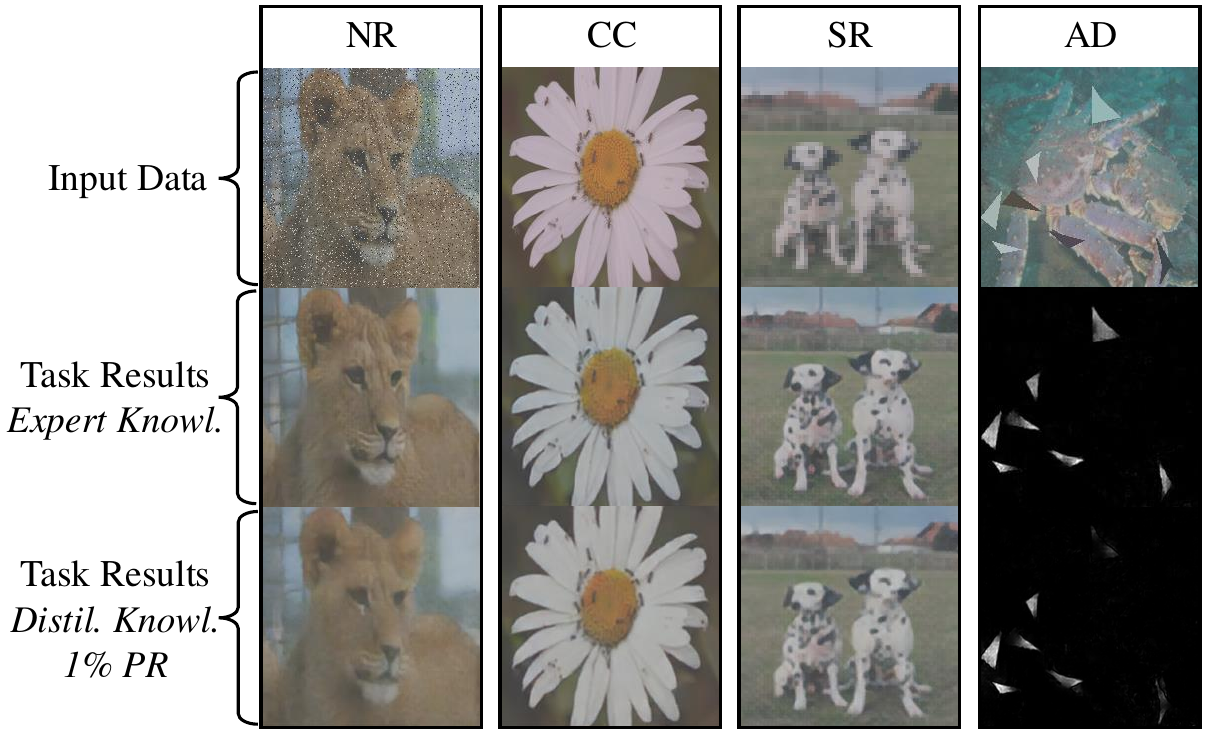}
    \caption{Comparison between the task results of full expert knowledge and of DK with 1\% PR.}
    \label{fig_eval_1}
\end{figure}

Fig.~\ref{fig_eval_1} compares the task results given full expert knowledge and DK of 1\% PR.
\rev{From the figure,} the DK is able to obtain output results that are very close to those of the full expert knowledge. 
\rev{Therefore, the} DeKAP effectively distills the expert knowledge into a significantly smaller set of parameters.

%% file: sections/eval_res/eval_2.tex
We then demonstrate the efficiency of the DeKAP in achieving KA compared with the existing approaches.
Specifically, we emulate the equalization-based approaches in~\cite{Sana23GC_Semantic} and~\cite{Huttebraucker24ISWCS_Soft} in the \emph{equal-latent} case, using multiple dense neural layers to learn the transformation between semantic latent spaces and sharing these layers for KA.
The \emph{data-adapt} case represents the adaptation-based approaches in~\cite{Zhang23JSAC_Deep,Si24TWC_Post}, where a set of semantic features of data and task results are shared to adapt the complete parameters.
The \emph{param-align} case is based on~\cite{Hu24Arxiv_Zero}, where the task adaptation is restricted to modifying sparse parameters, which are shared to achieve KA.

\fref{fig_exp2}{fig_exp2_a} shows the residual alignment loss versus the communication resources used for achieving the KA.
Here, the residual alignment loss is the ratio between the loss after and before the KA, averaged for the four tasks, and a unit communication resource is defined as the traffic load to transmit the DK of 1\% PR. 
As shown in \fref{fig_exp2}{fig_exp2_a}, DeKAP reduces the residual KA of the second best by 18\%.
\fref{fig_exp2}{fig_exp2_b} shows the residual alignment loss versus the computation resources.
A unit computation resource is defined as the number of float operations for optimizing the DK of 5\% PR for one epoch, and the same amount of training data is provided in all the cases.
As shown in \fref{fig_exp2}{fig_exp2_b}, DeKAP uses only 14\% of computation resources to achieve the performance \rev{same} as the second best.

\ifSingleColumn
    \renewcommand{\figwidth}{0.33\linewidth}
\else
    \renewcommand{\figwidth}{0.46\linewidth}
\fi

\begin{figure}[!t]
    \centering 
    \begin{subfigure}{\figwidth}
        \centering
        \includegraphics[width=\linewidth]{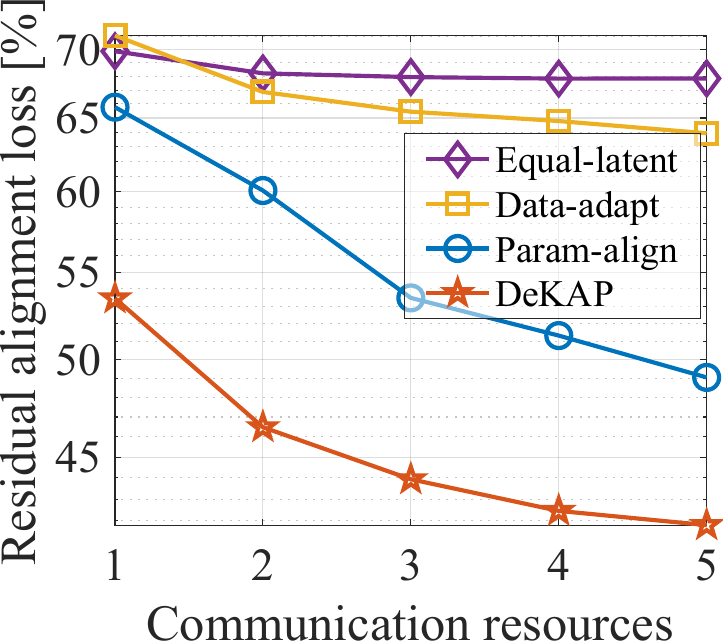}  
        \caption{}
        \label{fig_exp2_a}
    \end{subfigure}
    \ifSingleColumn
    \hspace{3ex}
    \else
    \hspace{1ex}
    \fi
    \begin{subfigure}{\figwidth}
        \centering
        \includegraphics[width=\linewidth]{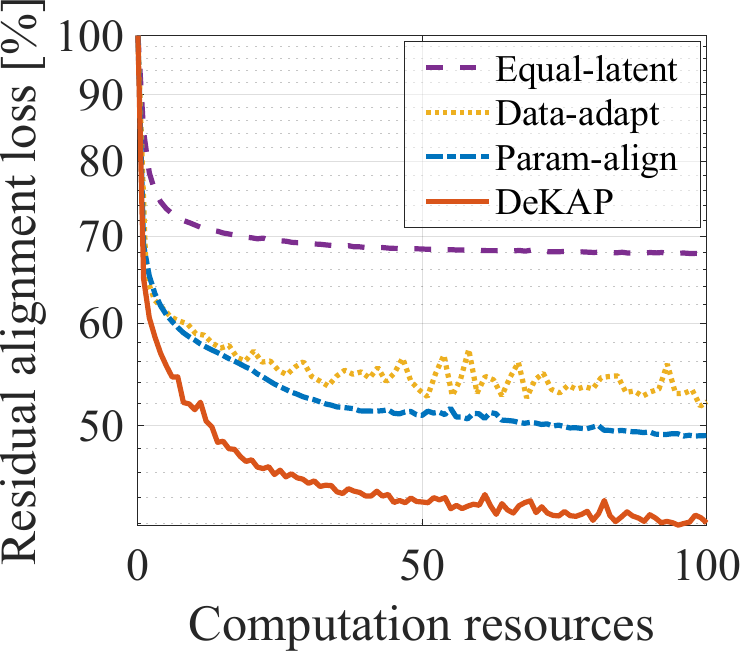}  
        \caption{}
        \label{fig_exp2_b}
    \end{subfigure}
    \caption{(a) Communication and (b) computation efficiency of the DeKAP compared to existing approaches for KA.}
    \label{fig_exp2}
\end{figure}

%% file: sections/eval_res/eval_3.tex
\ifSingleColumn
    \renewcommand{\figwidth}{0.33\linewidth}
\else
    \renewcommand{\figwidth}{0.46\linewidth}
\fi

\begin{figure}[!t]
    \centering 
    \begin{subfigure}{\figwidth}
        \centering
        \includegraphics[width=\linewidth]{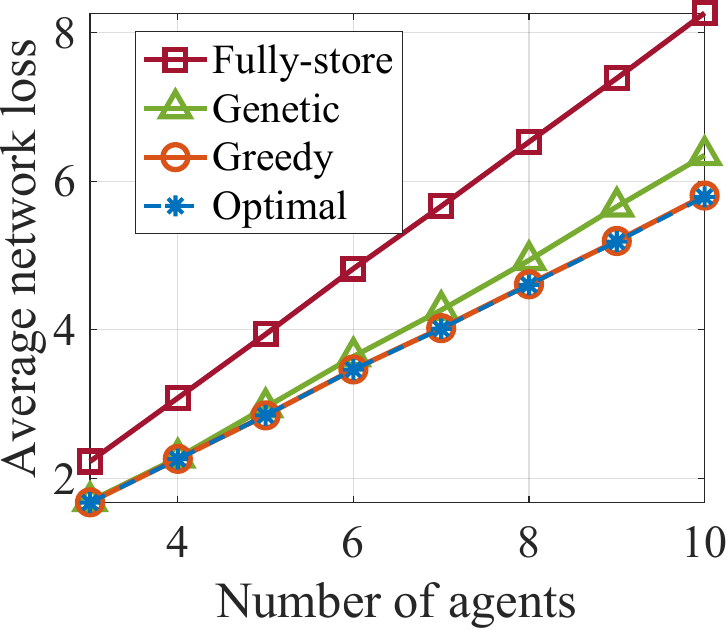}  
        \caption{}
        \label{fig_exp3_a}
    \end{subfigure}
    \ifSingleColumn
    \hspace{3ex}
    \else
    \hspace{1ex}
    \fi
    \begin{subfigure}{\figwidth}
        \centering
        \includegraphics[width=\linewidth]{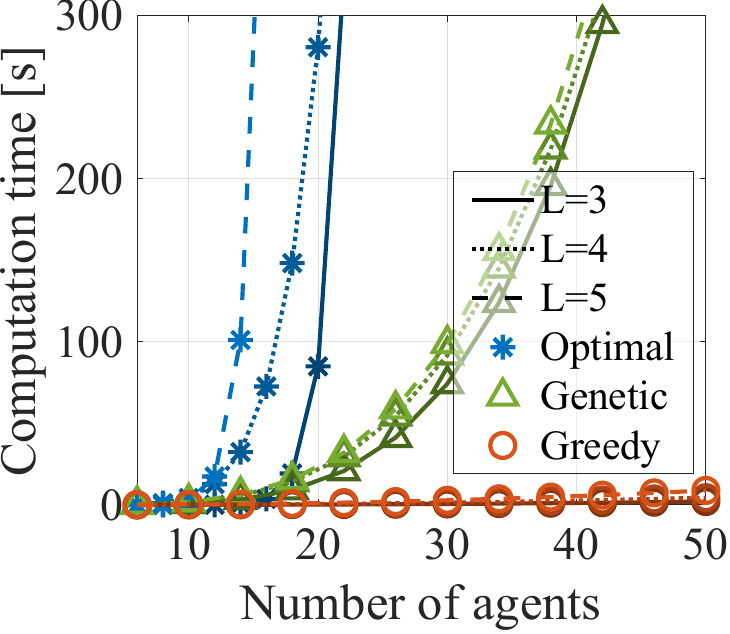}  
        \caption{}
        \label{fig_exp3_b}
    \end{subfigure}
    \caption{(a) Resulting average network loss and (b) computation time of different algorithms for the DK allocation. }
    \label{fig_exp3}
\end{figure} 

We then evaluate the multi-level DK allocation. 
In \fref{fig_exp3}{fig_exp3_a}, we sample 100 networks and compare the average network losses of different algorithms. 
\emph{Optimal} indicates solving the ILP using Gurobi~\cite{gurobi}.
\emph{Greedy} is the proposed greedy algorithm.
\emph{Genetic} is the genetic algorithm~(GA), a typical heuristic algorithm for integer programming~\cite{ga_alg}.
Moreover, in \rev{the \emph{fully-store}} case, every agent stores the complete multi-level DK, which is used as the baseline.
\fref{fig_exp3}{fig_exp3_a} shows that the local optimum found by the greedy algorithm is very close to the global optimum, reducing the average network loss by 28\% compared to the baseline.
In contrast, the performance of GA deteriorates quickly as $N$ increases. 
In \fref{fig_exp3}{fig_exp3_b}, we compare the computation time required by different algorithms for $L=3$, $4$, and $5$.
\rev{From the figure, the} computation time grows exponentially in the optimal and genetic cases. In comparison, the proposed greedy algorithm remains highly efficient for $N=50$, significantly boosting the DK allocation.

%% file: sections/6_conclusion.tex
\section{Conclusion}\label{SEC_CONCLU}

We have proposed \rev{the} DeKAP, a novel protocol to ensure aligned knowledge for SC in AI agent networks. 
By distilling the expert knowledge of agents into low-rank matrices, \rev{the} DeKAP generates multi-level DK and allocates to agents to minimize the alignment loss.
We have formulated an ILP for the optimization of DK allocation and proposed an efficient greedy algorithm.
Evaluation on image processing tasks validates the superior efficiency of \rev{the} DeKAP in minimizing the alignment loss.
Furthermore, the optimal allocation of DK reduces the network alignment loss, transmission overhead, and storage cost by 28\%, where the greedy algorithm achieves near global optimum \rev{but} significantly less computation time.

%% file: main.bbl
\begin{thebibliography}{10}
\providecommand{\url}[1]{#1}
\csname url@samestyle\endcsname
\providecommand{\newblock}{\relax}
\providecommand{\bibinfo}[2]{#2}
\providecommand{\BIBentrySTDinterwordspacing}{\spaceskip=0pt\relax}
\providecommand{\BIBentryALTinterwordstretchfactor}{4}
\providecommand{\BIBentryALTinterwordspacing}{\spaceskip=\fontdimen2\font plus
\BIBentryALTinterwordstretchfactor\fontdimen3\font minus
  \fontdimen4\font\relax}
\providecommand{\BIBforeignlanguage}[2]{{%
\expandafter\ifx\csname l@#1\endcsname\relax
\typeout{** WARNING: IEEEtran.bst: No hyphenation pattern has been}%
\typeout{** loaded for the language `#1'. Using the pattern for}%
\typeout{** the default language instead.}%
\else
\language=\csname l@#1\endcsname
\fi
#2}}
\providecommand{\BIBdecl}{\relax}
\BIBdecl

\bibitem{Chen20SUMMIT_From}
Y.~Chen, P.~Zhu, G.~He, X.~Yan, H.~Baligh, and J.~Wu, ``{From connected people,
  connected things, to connected intelligence},'' in \emph{Proc. 6G SUMMIT},
  Levi, Finland, Mar. 2020.

\bibitem{Xie21TSP_Deep}
H.~Xie, Z.~Qin, G.~Y. Li, and B.-H. Juang, ``Deep learning enabled semantic
  communication systems,'' \emph{IEEE Trans. Signal Process.}, vol.~69, pp.
  2663--2675, Apr. 2021.

\bibitem{Qin24PIEEE_AI}
Z.~Qin, L.~Liang, Z.~Wang, S.~Jin, X.~Tao, W.~Tong, and G.~Y. Li, ``{AI
  empowered wireless communications: From bits to semantics},'' \emph{Proc.
  IEEE}, vol. 112, no.~7, pp. 621--652, Jul. 2024.

\bibitem{Bourtsoulatze19TCCN_Deep}
E.~Bourtsoulatze, D.~Burth~Kurka, and D.~Gündüz, ``Deep joint source-channel
  coding for wireless image transmission,'' \emph{IEEE Trans. Cogn. Commun.
  Netw.}, vol.~5, no.~3, pp. 567--579, Sep. 2019.

\bibitem{Rossi22Huawei_Network}
D.~Rossi and L.~Zhang, ``Network artificial intelligence, fast and slow,'' in
  \emph{Proc. Int. Workshop Native Netw. Intell.}, Rome, Italy, Dec. 2022.

\bibitem{Zhang23JSAC_Deep}
H.~Zhang, S.~Shao, M.~Tao, X.~Bi, and K.~B. Letaief, ``Deep learning-enabled
  semantic communication systems with task-unaware transmitter and dynamic
  data,'' \emph{IEEE J. Sel. Areas Commun.}, vol.~41, no.~1, pp. 170--185, Jan.
  2023.

\bibitem{Si24TWC_Post}
P.~Si, R.~Liu, L.~Qian, J.~Zhao, and K.-Y. Lam, ``Post-deployment fine-tunable
  semantic communication,'' \emph{IEEE Trans. Wireless Commun.}, vol.~24,
  no.~1, pp. 35--50, Jan. 2025.

\bibitem{Choi24TVT_Semantics}
J.~Choi, J.~Park, S.-W. Ko, J.~Choi, M.~Bennis, and S.-L. Kim, ``Semantics
  alignment via split learning for resilient multi-user semantic
  communication,'' \emph{IEEE Trans. Veh. Technol.}, vol.~73, no.~10, pp.
  15\,815--15\,819, Oct. 2024.

\bibitem{Sana23GC_Semantic}
M.~Sana and E.~C. Strinati, ``Semantic channel equalizer: Modelling language
  mismatch in multi-user semantic communications,'' in \emph{Proc. IEEE
  GLOBECOM}, Kuala Lumpur, Malaysia, Dec. 2023.

\bibitem{Fiorellino24Arxiv_Dynamic}
S.~Fiorellino, C.~Battiloro, E.~C. Strinati, and P.~Di~Lorenzo, ``Dynamic
  relative representations for goal-oriented semantic communications,'' in
  \emph{Proc. EUSIPCO}, Lyon, France, Aug. 2024.

\bibitem{Hu24Arxiv_Zero}
J.~Hu and G.~Y. Li, ``{Zero-forget preservation of semantic communication
  alignment in distributed AI networks},'' \emph{arXiv:2411.19385}, Nov. 2024.

\bibitem{Hu22ICLR_LoRA}
E.~J. Hu, Y.~Shen, P.~Wallis, Z.~Allen-Zhu, Y.~Li, S.~Wang, L.~Wang, and
  W.~Chen, ``{LoRA: Low-rank adaptation of large language models},'' in
  \emph{Proc. ICLR}, Online, Apr. 2022.

\bibitem{zhang2021self}
L.~Zhang, C.~Bao, and K.~Ma, ``Self-distillation: Towards efficient and compact
  neural networks,'' \emph{IEEE Trans. Pattern Anal. Mach. Intell.}, vol.~44,
  no.~8, pp. 4388--4403, Aug. 2021.

\bibitem{gurobi}
{Gurobi Optimization, LLC}, ``Gurobi optimizer,'' 2024, available:
  \url{https://www.gurobi.com}.

\bibitem{Van17NIPS_Neural}
A.~Van Den~Oord, O.~Vinyals \emph{et~al.}, ``Neural discrete representation
  learning,'' in \emph{Proc. NeurIPS}, Long Beach, CA, USA, Dec. 2017.

\bibitem{vgg16loss}
J.~Johnson, A.~Alahi, and L.~Fei-Fei, ``{Perceptual losses for real-time style
  transfer and super-resolution},'' in \emph{Proc. ECCV}, Amsterdam, The
  Netherlands, Oct. 2016.

\bibitem{Huttebraucker24ISWCS_Soft}
T.~H{\"u}ttebr{\"a}ucker, M.~Sana, and E.~C. Strinati, ``Soft partitioning of
  latent space for semantic channel equalization,'' in \emph{Proc. IEEE ISWCS},
  Rio de Janeiro, Brazil, Jul. 2024.

\bibitem{ga_alg}
K.~Deep, K.~P. Singh, M.~Kansal, and C.~Mohan, ``A real coded genetic algorithm
  for solving integer and mixed integer optimization problems,'' \emph{Appl.
  Math. Comput.}, vol. 212, no.~2, pp. 505--518, Jun. 2009.

\end{thebibliography}
